# An extended empirical model for L- and M-shell ionizations of atoms

M. R. Talukder
Department of Applied Physics & Electronic Engineering
University of Rajshahi, Rajshahi-6205, Bangladesh.

**Abstract**

An extension of the analytical model of Talukder et al (Int. J. Mass Spectrom. 269 (2008) 118) is proposed to estimate electron impact single L- and M-shell ionization cross sections of atoms with incident energy from threshold to ultra-relativistic range. Comparisons are made with other theoretical calculations. It is found that this model agrees well with the experimental data and quantum calculations.

**1. Introduction**

Bell et al. [1] proposed an empirical model (BELL) for the calculation of electron impact ionization cross sections (EIICS) for the limited range of incident energies and a few atomic species with six fitting parameters. Talukder et al. [2] simplified the BELL (SBELL) model incorporating the ionic and relativistic factors in it, with only two fitting parameters instead of six, in the relativistic incident energies for the fast estimation of EIICS of K-shell applicable for targets with atomic number Z = 1–92. The parameters of the SBELL model, which have been generalized in terms of the ionization potentials and are independent of species, are found successful for the description of EIICS data for K-shell. To the best of our knowledge, the SBELL model is a good performer in describing K-shell EIICS data. So it is intended to extend the SBELL model, because of its level of performance, simplicity and fast generation of cross section data as needed for modelling, for the estimation of EIICS for L- and M-shells in the ultra-relativistic incident energies for atomic species.

**2. Outline of the model**

The SBELL model is given [2] by

$$\sigma_{BR}(E) = \left\{ A \ln\left(\frac{E}{I_j}\right) + B\left(1 - \frac{I_j}{E}\right) \right\} a_o^2 , \qquad (1)$$

where $I_j$, $E$, $a_o = 0.529 \times 10^{-8}$ cm, $A$ and $B$ are the ionization potentials of *L*- and *M*-shell of the targets, energy of the incident electron, first Bohr radius, and fitting parameters, respectively. The Gryzinski's relativistic term [2] is given by

$$F_{SG} = \frac{2(1+J/U)^2}{J(1+2J/U)}, \qquad (2)$$

where $J = mc^2/I_j$, $m$ is the mass of electron, $c$ is the velocity of light in vacuum, and $U = E/I_j$ is the reduced energy. The ionic correction factor [2] $F_I$ is given by

$$F_{IM} = \left\{1 + n\left(\frac{q}{ZU}\right)\right\}^\lambda, \qquad (3)$$

where $n$ and $\lambda$ are the fitting parameters, ion of charge $q = Z - N_j$, $Z$ is the atomic number, and $N_j$ is the number of electrons in the orbit considered. The values of $n = 3.65$ and $\lambda = 1.15$ are used [2] in the calculation. The SBELL model for the electron impact $L$- and $M$-shell ionization cross section $\sigma_{SBELL}$ is given [2] by

$$\sigma_{SBELL} = N_j F_{IM} F_{SG} \sigma_{BR}(E). \qquad (4)$$

In Eq. (4) the fitting parameters $A$ and $B$ are generalized by making them dependent on $I_j$. Ionization potential is normalized by $U_R = I_j/R$, where $R$ is the Rydberg energy. The parameters $A$ and $B$ are then expressed by

(a) $L$-shell: (i) for $s$ orbit

$$A = \frac{4850.99 U_R}{(1+5.306 U_R)^{3.15}}, \qquad (5)$$

$$B = -\frac{4849.48 U_R}{(1+1.530 U_R)^{3.18}}, \qquad (6)$$

(ii) for $p$ orbit

$$A = \frac{4553.23 U_R}{(1+7.73 U_R)^{3.05}}, \qquad (7)$$

$$B = -\frac{172.94 U_R}{(1+1.91 U_R)^{3.19}}, \qquad (8)$$

(b) $M$-shell: (i) for $s$ orbit

$$A = \frac{496.47 U_R}{(1+1.57 U_R)^{3.38}}, \qquad (9)$$

$$B = -\frac{391.15 U_R}{(1+1.40 U_R)^{3.38}}, \qquad (11)$$

(ii) for $p$ orbit

$$A = \frac{2455.66 U_R}{(1+4.97 U_R)^{3.1}}, \qquad (12)$$

$$B = -\frac{75335.46 U_R}{(1+17.50 U_R)^{3.1}}, \qquad (13)$$

(iii) for $d$ orbit

$$A = \frac{2.53 U_R}{(1 + 0.82 U_R)^{2.9}}, \quad (14)$$

$$B = -\frac{10.04 U_R}{(1 + 1.63 U_R)^{2.9}}. \quad (15)$$

**3. Results and discussions**

The ionization potentials are taken from Desclaux [3]. In the analysis, 18 atomic targets Ar, Ni, Cu, Sr, Y, Pd, Ag, In, Sn, Ba, Sm, Ho, Yb, Ta, Au, Pb, Bi, and U for *L*-shell, 4 targets Au, Pb, Bi, and U for *M*-shell, whose experimental data are available have been considered. A few will be presented in the short note for example. The same fitting procedure is applied as used in [2].

Experimental and quantum EIICS data for L-shell, are taken from Ishii et al [4], Hoffmann et al [5], Khare et al (plane wave Born approximation (PWBA)) [6] and Scofield et al (relativistic PWBA (RPWBA)) [7], respectively, are displayed in Fig.1. Empirical model, MBELL [8], calculations have been included to judge the performance of the SBELL model. The predictions of SBELL model agree well with the experimental data and quantum calculations as shown in Fig. 1. The MBELL model underscores the experimental as well as quantum data. So it is evidenced that SBELL model provides better performance for the calculation of EIICS with respect to the MBELL model.

Experimental and quantum EIICS data for *M*-shell, are collected from [4], [5], and [6], and are depicted in Fig. 2. PWBA cross sections overestimate the experimental data over the whole range of incident energies. But, the predictions of SBELL model agree well with the experimental data.

**4. Conclusions**

The work reports the performance of the SBELL model for the calculation of EIICS for L- and M-shell atoms for the incident energies from threshold to ultra-relativistic range. SBELL model provides better performance as compared to the MBELL model.

**Acknowledgments**

This work was done within the framework of the Associateship Scheme of the Abdus Salam International Centre for Theoretical Physics (ICTP), Trieste, Italy. Author would like to thank ICTP for kind hospitality and support.

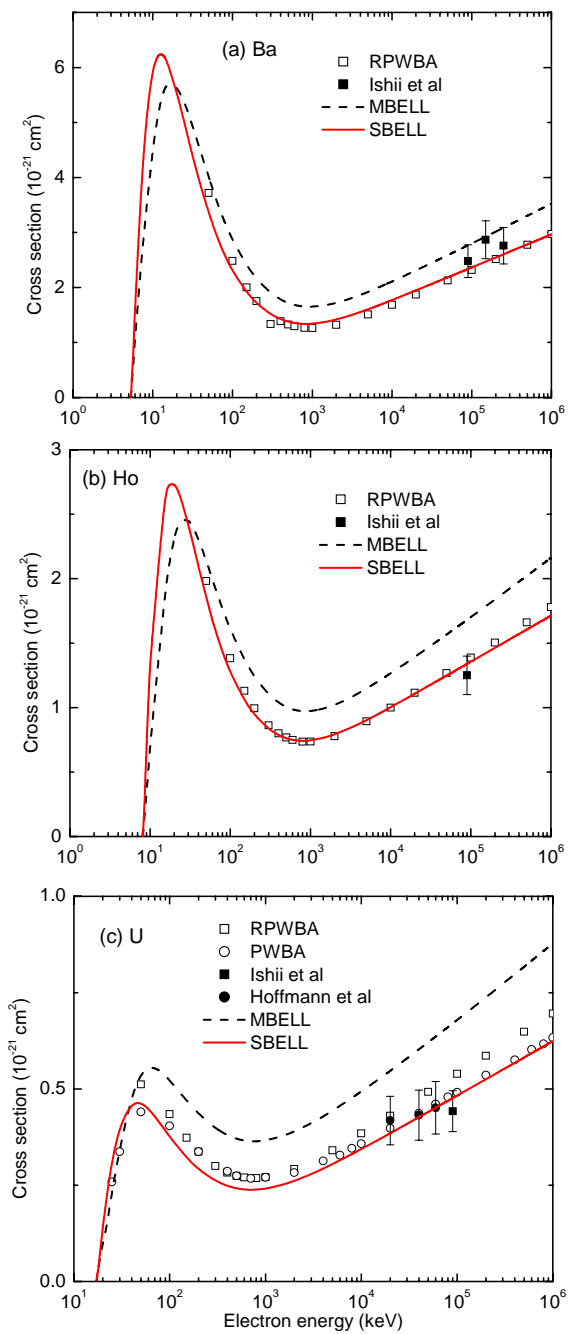

FIG. 1. Electron impact ionization cross sections of *L*-shell for: a) Ba, b) Ho, and c) U.

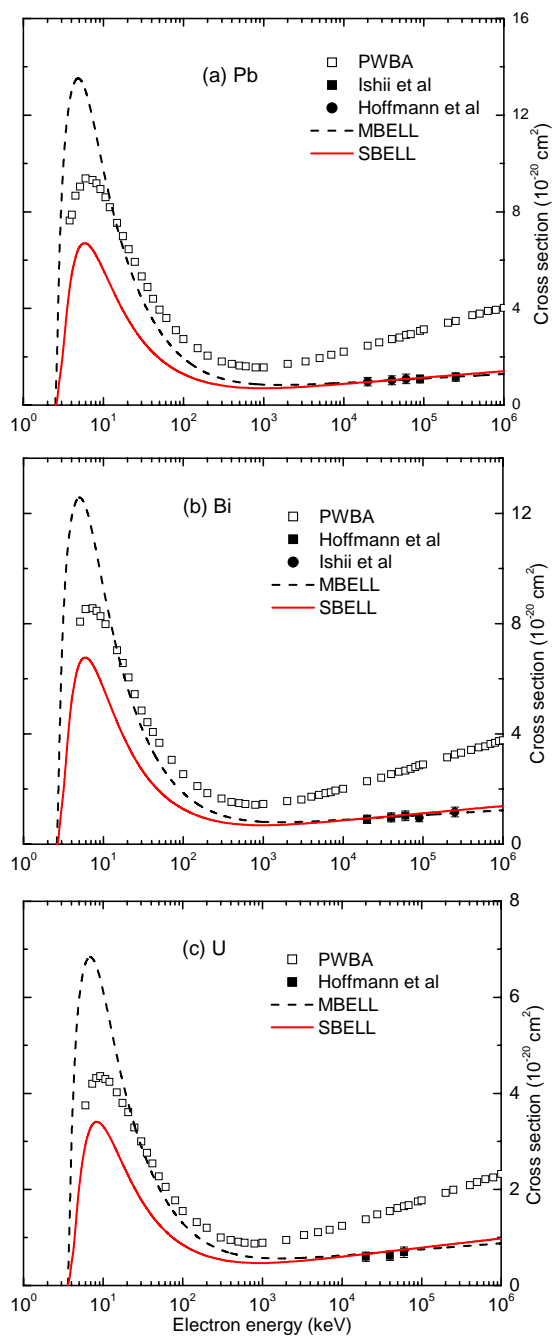

FIG. 2. Electron impact ionization cross sections of *M*-shell for: a) Pb, b) Bi, d) U.